\documentclass[twoside]{dis08}
\usepackage[latin1]{inputenc}
\usepackage[dvips]{graphicx,epsfig,color}
\usepackage{wrapfig,rotating}
\usepackage{amssymb,amsmath,array}

\begin{document}
\title{About Geometric Scaling and Small-\boldmath{$x$} Evolution 
at RHIC and LHC}

\author{Andre Utermann
%
%
\vspace{.3cm}\\
%
Department of Physics and Astronomy,
Vrije Universiteit Amsterdam, \\
De Boelelaan 1081, 1081 HV Amsterdam, The Netherlands
%
}

\maketitle

\begin{abstract}
  We show that the whole range of RHIC data for hadron production in
  $d$-$Au$ collisions is compatible with geometric scaling.  To
  establish the scaling violations expected from small-$x$ evolution a
  larger kinematic range in transverse momentum and rapidity would be
  needed.  We point out that the fall-off of the $p_t$ distribution of
  produced hadrons at large $p_t$ is a sensitive probe of small-$x$
  evolution especially at the LHC.
\end{abstract}

It is clearly observed that the small-$x$ DIS data show the property
of geometric scaling (GS) \cite{Stasto:2000er}, i.e.\ the cross
section depends on the combination $Q^2/Q_s^2(x)$ only, where $Q_s(x)$
is referred to as the saturation scale.  On the other hand, geometric
scaling is a feature of the asymptotic solutions of nonlinear
evolution equations, such as the BK equation \cite{BK}. Hence,
geometric scaling is seen as a strong indication for small-$x$ gluon
saturation.

A phenomenological study of DIS data using a model for the dipole
cross section was performed by Golec-Biernat and W\"{u}sthoff
(GBW)~\cite{GBW}. They found that the inclusive HERA data at low
$x\lesssim 0.01$ could be described well by a dipole cross section
$\sigma = \sigma_0 N_{\rm GBW}(r_t,x)$, where $\sigma_0 \simeq
23\;{\rm mb}$ and the scattering amplitude $N_{\rm GBW}$ is given by
\begin{equation}
N_{\rm GBW}(r_t,x) = 1-\exp\left(-\tfrac{1}{4} r_t^2
Q_s^2(x) \right). 
\label{NGBW}
\end{equation}
 The $x$-dependence of the saturation scale is given by
\begin{equation}
Q_s(x) = Q_0\,
\left(\frac{x_0}{x}\right)^{\lambda/2}, \,\quad Q_0=1\,\mathrm{GeV}\,
\label{Qsx2}
\end{equation}
where the parameters $x_0 \simeq 3 \times 10^{-4}$ and $\lambda \simeq
0.3$ are fitted to the small-$x$ data. The amplitude (\ref{NGBW})
depends on $x$ and $r_t$ (the transverse size of the dipole) only
through the combination $r_t^2\,Q_s^2(x)$. In DIS, this directly
results in a scaling cross section. Hence, in the dipole picture,
GS is equivalent to a dependence of the amplitude on
$r_t^2\,Q_s^2(x)$ only.

Despite that GS is expected already sub-asymptotically in a growing
range around $Q_s$ \cite{MTIIM}, the values of $x$ probed at recent colliders may
be not small enough to expect GS and it may be more important to
test violations of GS to establish small-$x$ evolution.  In order to
investigate GS violations in the RHIC data, in \cite{DHJ1,DHJ2}
a phenomenological model has been put forward.
We will refer to this model as the DHJ model. It offers a good
description of the $p_t$ distribution of hadrons produced in $d$-$Au$
collisions at RHIC in the forward region \footnote{As it turned out
  the study of Ref.\ \cite{DHJ2} contained an artificial upper limit
  on the $x_1$ integration to exclude large $x_2$. Without this cut,
  the larger $p_t$ data for $y_h=0, 1$ are in fact not well-described
  by the DHJ model.}.
 
According to Refs.\ \cite{DHJ1,DHJ2} the cross section of hadron
production in high-energy nucleon-nucleus collisions can be described in
terms of the dipole scattering amplitude,
\begin{eqnarray}
\frac{ d\,N_h}{d\,y_h\: d^2p_t} &=& 
{K(y_h) \over (2\pi)^2} \int_{x_F}^{1} d\,x_1 \; {x_1\over x_F}
\Bigg[\sum_q f_{q/p}(x_1,p_t^2)\, N_F \left({x_1\over x_F}p_t,x_2\right)\,
D_{h/q}\left({x_F\over x_1},p_t^2\right)
\nonumber \\
& &+~
f_{g/p}(x_1,p_t^2)\, N_A \left({x_1\over x_F}p_t,x_2\right)\, 
D_{h/g}\left({x_F\over x_1},p_t^2\right)\Bigg]~.
\label{eq:conv2}
\end{eqnarray}
Here $N_F$ describes
a quark scattering off the nucleus, while $N_A$ applies to a gluon.
The parton distribution functions $f_{q/p}$ and the fragmentation
functions $D_{h/q}$ are considered at the scale $Q^2=p_t^2$, which we
will always take to be larger than 1 GeV$^2$.  The momentum fraction
of the target partons equals $x_2=x_1\exp(-2y_h)$. We can
to good approximation neglect finite mass effects, i.e.\ we equate the
pseudorapidity $\eta$ and the rapidity $y_h$ and use
$x_F=\sqrt{p_t^2+m^2}/\sqrt{s}\exp(\eta)\approx
p_t/\sqrt{s}\exp(y_h)$.  Finally, there is an overall $y_h$ dependent
$K$-factor that effectively accounts for NLO corrections.  It should be noted that
the scaling properties of the dipole scattering amplitude are not
directly visible in the hadron production data, due to its
convolution
with the parton distribution and fragmentation functions.

The dipole scattering amplitude of the DHJ model is
given by~\cite{DHJ1,DHJ2}:
\begin{equation} 
N_A({q}_t,x_2)  \equiv \int d^2 r_t\: e^{i \vec{q}_t
    \cdot \vec{r}_t} \left[1-\exp\left(-\tfrac{1}{4}(r_t^2
      Q_s^2(x_2))^{\gamma(q_t,x_2)}\right) \right]~.
\label{NA_param}
\end{equation}
Note that $\gamma$ is a function of $q_t$ rather than $r_t$. This allows one to
compute the Fourier transform more easily. 
The corresponding expression $N_F$ for quarks is obtained from $N_A$
by the replacement $(r_t^2Q_s^2)^\gamma \to ((C_F/C_A) r_t^2Q_s^2)^\gamma$,  
with $C_F/C_A=4/9$.
The exponent $\gamma$ is usually referred to as the
``anomalous dimension'', although the connection between $N_{A/F}$ and
the gluon distribution inside the nucleus cannot always be made.

The anomalous dimension of the DHJ model is parameterized as
\begin{equation}
\gamma_{\rm DHJ}(q_t,x_2) =
\gamma_s + (1-\gamma_s)\, \frac{|\log(q_t^2/Q_s^2(x_2))|}{\lambda
y+d\sqrt{y}+|\log(q_t^2/Q_s^2(x_2))|}, \label{gammaparam}
\end{equation}
where $y=\log 1/x_2$ is minus the rapidity of the target parton. The
saturation scale $Q_s(x_2)$ and $\lambda$ are taken from
the GBW model (\ref{Qsx2}). Here $Q_s$ includes a
larger value of $Q_0\approx 1.63\;{\rm GeV}$ to account for the size
of the nucleus. The parameter $d=1.2$ was fitted to the data.  This
choice of $\gamma$ leads to a geometric scaling solution at $q_t=Q_s$
where $\gamma=\gamma_s=0.628$ and incorporates to a certain extent the
violation expected from BFKL evolution for larger $q_t$, see
\cite{MTIIM}.  However, an analysis of the BK equation suggests that a
smaller value of $\gamma\approx 0.44$ \cite{BUW} may be more
appropriate at $Q_s$ where the BFKL equation does not apply.  

Not only a constant $\gamma$ would lead to GS but also a $\gamma$ that
depends on $q_t^2/Q_s^2$ or $r_t^2 Q_s^2$ only. To check explicitly
whether the RHIC data are compatible with GS, we propose a new scaling
parameterization of $\gamma$ that is similar in form to that of the
DHJ model, but does not have the GS violating behavior nor the
logarithmic rise expected from the BFKL (and more generally, BK)
equation. The parameterization that we adopted reads
\begin{equation}
 \gamma_{\rm GS}(w=q_t/Q_s)=\gamma_1+(1-\gamma_1)\frac{(w^a-1)}{(w^a-1)+b}\,.
\label{gamma_alpha}
\end{equation}
The two free parameters $a$ and $b$ will be fitted to the RHIC data.
There are two major differences between the chosen parameterization
$\gamma_{\rm GS}$ (\ref{gamma_alpha}) and $\gamma_{\rm DHJ}$
(\ref{gammaparam}). Firstly, $\gamma_{\rm GS}$ does not depend on the
rapidity $y$ explicitly. Therefore the resulting dipole scattering amplitude
respects geometric scaling.  Secondly, $\gamma_{\rm GS}$ approaches
the large $q_t$ limit of 1 much faster.  This will lead to different
large momentum slopes of the amplitude (\ref{NA_param}) and therefore
to different predictions for the large $p_t$ slope using Eq.\
(\ref{eq:conv2}). For large $w$ the exponential function can be
expanded and the fall-off of the dipole scattering amplitude
(\ref{NA_param}) is given by \cite{buw2},
\begin{equation}
  N_A({q}_t) \stackrel{q_t \gg Q_s}{\propto}
\left\{\begin{array}{cl}
    \frac{Q_s^2}{q_t^4\log(q_t^2/Q_s^2)}& \mbox{} \quad 
\text{for $\gamma$ of Eq.\ (\ref{gammaparam})}
\\[2ex]
\frac{Q_s^{2+a}}{q_t^{4+a}}& \mbox{} \quad \text{for $\gamma$
      of Eq.\ (\ref{gamma_alpha})} 
\end{array}\right.\,.
\label{NA_asympt}
\end{equation}
Empirically, we find that the $p_t$ distribution falls off even faster.
We note that the fall-off with $p_t$ is not determined by
the size of the scaling violations. In order to observe such
violations, one has to study both the $y_h$ and $p_t$ dependence over a
significantly large range. 
Let us mention explicitly that our parameterization is not meant to
replace other, physically better motivated models but to investigate
in a general way which conclusion can really be drawn unambiguously
from the RHIC data in the central and forward regions.

\begin{figure}[htb]
\centering
\includegraphics*[width=69mm]{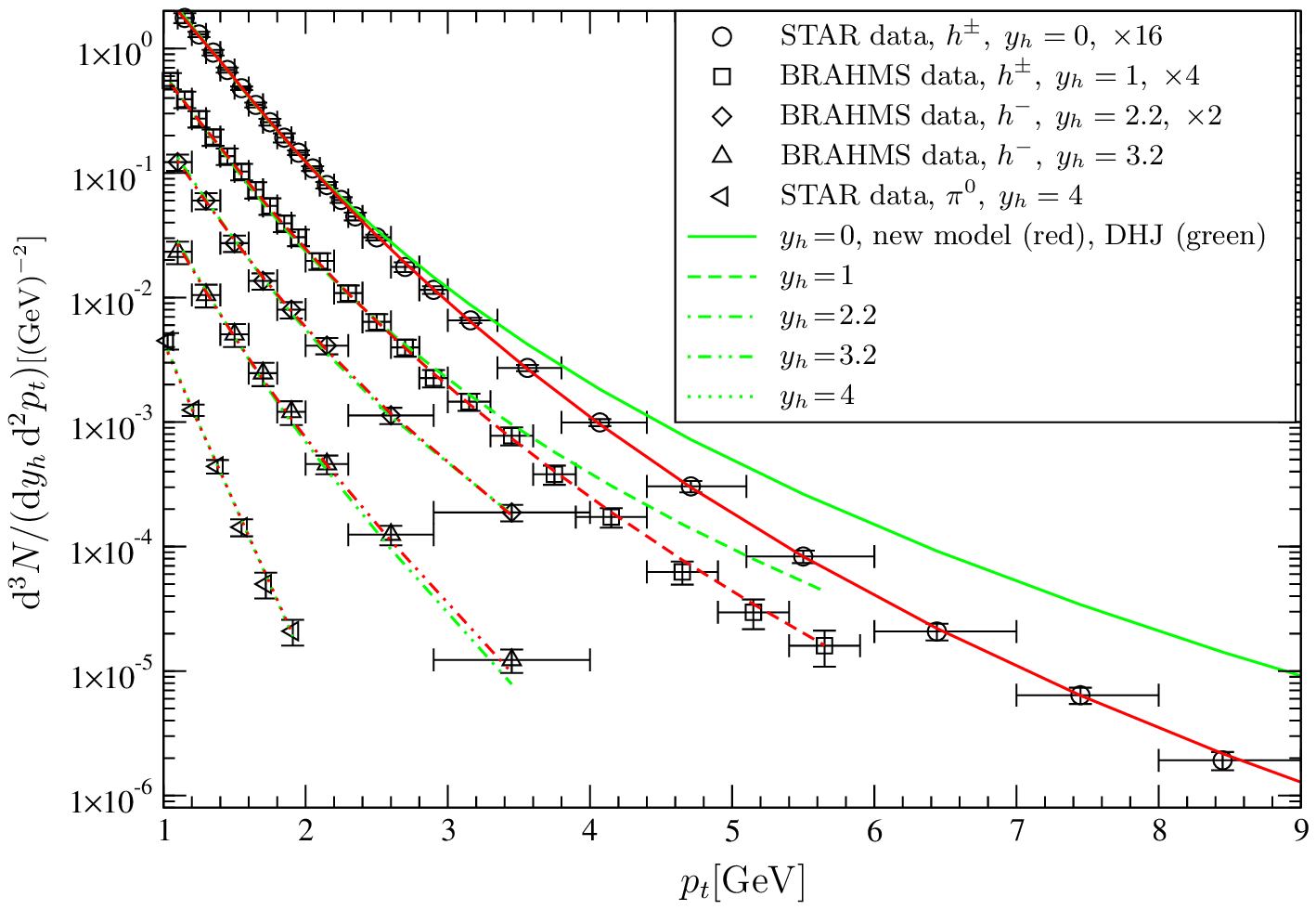}
\includegraphics*[width=69mm]{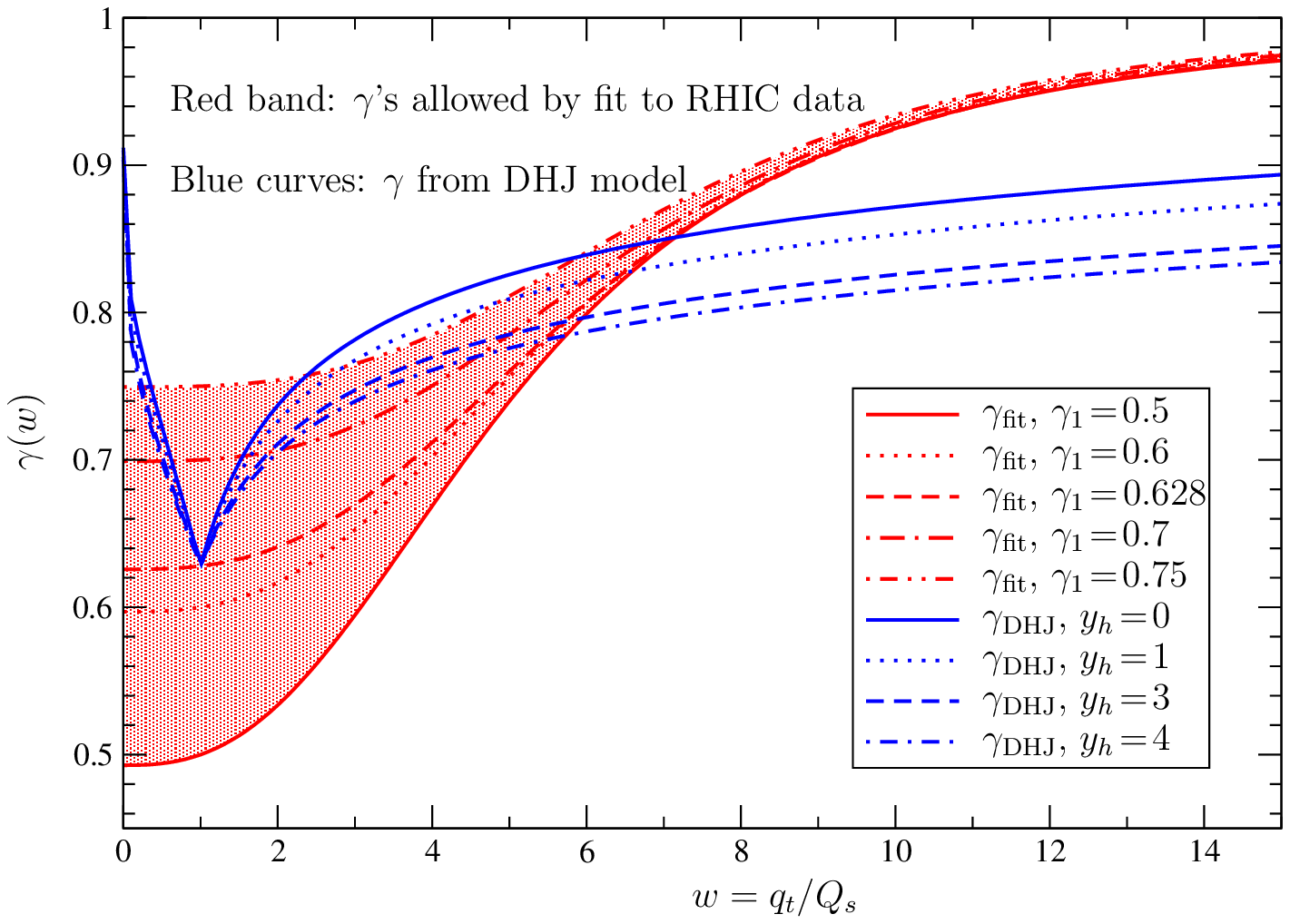}
\caption{ \small\label{fig_RHIC} a) Transverse momentum distribution
  of produced hadrons in $d$-$Au$ collisions as measured at RHIC
  (symbols) for various rapidities $y_h$.  To make the plot clearer,
  the data and the curves for $y_h=0, 1$ and $2.2$ are multiplied with
  arbitrary factors, namely 16, 4 and 2, respectively.  \mbox{b)
    Various} fits of $\gamma_{\rm GS}(w)$, which describe the RHIC
  data equally well. For comparison we show curves representing
  $\gamma_{\rm DHJ}(w,y(w,y_h))$ at different rapidities $y_h$.}
\end{figure}

In Fig.~\ref{fig_RHIC} a) we show our estimate for $dN_h/(dy_h
d^2p_t)$ that follows from the integral in Eq.\ (\ref{eq:conv2}) by using
$\gamma_{\rm GS}(w)$ (\ref{gamma_alpha}). All $p_t$ distributions of
produced hadrons measured at RHIC in $d$-$Au$ collisions are well
described. At the saturation scale we have chosen here for
$\gamma_{\rm GS}(w=1)=\gamma_1$
the same value $\gamma_s=0.628$ as in the DHJ model.  We also take the
same parameterization of $Q_s(x)$.  We obtain the best fit of the data
for $a=2.82$ and $b=168$.  As mentioned, this LO analysis requires the
inclusion of a $K$-factor to account for NLO corrections, which are
expected to become more relevant towards central rapidity.  The $p_t$-independent $K$
factors we obtain for $y_h=0,1,2.2,3.2,4$ are for our model equal
to $K=3.4, 2.9, 2.0,1.6, 0.7$ and for the DHJ model $K=4.3, 3.3, 2.3,
1.7, 0.7$. For further details of the calculation see \cite{buw2}.

\mbox{From} this analysis we can conclude that a GS dipole 
amplitude is completely compatible with the data and therefore the
conclusion that GS violations are observed at RHIC cannot be drawn.
Of course, a scaling violating amplitude, i.e.\ a $\gamma$ that depends
on $w$ and the rapidity $y$ explicitly, is not ruled out by the data
either.  What can be concluded further is that the logarithmic rise of
$\gamma$ resulting from the BFKL evolution incorporated in the DHJ
model is ruled out in the central region, see Fig.~\ref{fig_RHIC} a).
However, where the DHJ model starts to deviate from the data $x_2$
becomes larger than 0.01, although $Q_s$ then is still larger than in DIS
at $x=0.01$. If one were to exclude the central rapidity RHIC data in
the model fit, one could also obtain a scaling model with a
logarithmically rising, or even constant, $\gamma$.

To indicate how much $\gamma$ is constrained by the RHIC data,
Fig.~\ref{fig_RHIC} b) shows various $\gamma_{\rm GS}(w)$'s that
describe the available data equally well.  They are all parameterized
as in Eq.\ (\ref{gamma_alpha}) with different $a$ and $b$ values.
Furthermore, we added lines representing $\gamma_{\rm DHJ}$
(\ref{gammaparam}) for different rapidities. For this one needs to
express the rapidity of the target parton $y$ in terms of $w$ and $y_h$,
see \cite{buw2} for details. It should be mentioned that the region
below $Q_s$, where the parameterization of $\gamma_{\rm DHJ}$ is not
smooth, is hardly probed at RHIC. Clearly, $\gamma$ is less well
determined close to the saturation scale than in the dilute region.
This is because the integrand entering the dipole scattering amplitude
(\ref{NA_param}) is only weakly dependent on $\gamma$ around the
saturation scale $r=1/Q_s$.  In addition, the forward data ($y_h=3.2$
and 4) are essentially sensitive only to $\gamma_1$, since they probe
the region where $w$ is close to 1.  Therefore, the rise of $\gamma$
with $w$ is effectively constrained only by the data for $y_h=0,1$. It
is apparent from Fig.~\ref{fig_RHIC} b) that the DHJ model fails for
larger $w$, i.e.\ larger $p_t$, in the central region but not in the
forward region where the probed values of $w$ are below 2.

Where the DHJ model curves deviate from the RHIC data, the probed
$x_2$-values are not very small. However, at LHC, due to the higher
energies, the region of small $x_2$ extends to a much larger range of
$p_t$, so that the predictions of the DHJ model and the new one will
be different even at small $x_2$.  It has to be mentioned that the
calculation of the $p_t$ distribution using our scaling model should
not be seen as a physically motivated prediction. However, a
comparison of the estimate using $\gamma_{\rm GS}$ fitted to RHIC data
with the estimate from the DHJ model allows drawing conclusions
anyway.  As we will explain, the estimates are so different that the
LHC should be able to rule out one of these models at much smaller
$x$.  Hence, the LHC can answer the question why the DHJ model fails
in the central region at RHIC. Either because the probed $x$ values
are not small enough or because some expectations from small-$x$
evolution has to be modified. For smaller $p_t$ the predictions of the
two models are expected to be comparable since only the region of
small values of $w$ is probed where $\gamma_{\rm DHJ}$ and
$\gamma_{\rm GS}$ are similar, see Fig.~\ref{fig_RHIC} b).  In the
very forward region, i.e.\ $y_h \approx 6 - 8$, only this region is
tested.  However, there is quite a large range where the probed values
of $x_2$ are small but the predictions are clearly different.  One
expects (see Fig.~\ref{fig_RHIC} b)) differences in the central region
between the two models for values of $w\gtrsim 3$, i.e.\ at RHIC for
$p_t\gtrsim 2.5\;{\rm GeV}$. Since $Q_s$ is larger at LHC such
differences show up at larger momenta around 5 GeV.  Even in the
central region such momenta imply values of $x_2$ smaller than 0.001.
Hence, a measurement of the slopes at moderate rapidities $y_h$ at LHC
would allow a discrimination between the two models in a region where
small-$x$ physics is expected to be applicable.  The slower fall-off
of the $p_t$ distribution in the DHJ model is a direct consequence of
the logarithmic rise of $\gamma$ towards 1. Since such a behavior of
$\gamma$ is a generic signature of BFKL evolution, these measurements
offer the possibility of testing whether such small-$x$ evolution is
actually relevant at present-day hadron colliders. Estimates of the
$p_t$ distributions for hadron production at LHC using both models can
be found in \cite{buw2}.
 
\begin{footnotesize}



%

\end{footnotesize}


\end{document}